\documentclass[preprint,showpacs,preprintnumbers,amsmath,amssymb]{revtex4}

\usepackage{graphicx}
\usepackage{epstopdf}
\usepackage{dcolumn}
\usepackage{bm}

\begin{document}

\preprint{}

\title{Plasmonic Waves on a Chain of Metallic Nanoparticles:  Effects of a Liquid Crystalline Host or an Applied Magnetic Field}  
\author{N.\ A.\ Pike and D.\  Stroud}
\affiliation{%
Department of Physics, Ohio State University, Columbus, OH 43210\\
}

\date{\today}

\begin{abstract}

A chain of metallic particles, of sufficiently small diameter and spacing, allows linearly polarized plasmonic waves to propagate along the chain.
In this paper, we consider how these waves are altered by an anisotropic host (such as a nematic liquid crystal) or an applied magnetic field. In a liquid crystalline host, with principal axis (director) oriented either parallel or perpendicular to the chain, we find that the dispersion relations of both the longitudinal ($L$) and transverse ($T$) modes are significantly altered relative to those of an isotropic host.  Furthermore, when the director is perpendicular to the chain, the doubly degenerate $T$ branch is split by the anisotropy of the host material.   With an applied magnetic  field ${\bf B}$ parallel to the chain, the propagating transverse modes are circularly polarized, and the left and right circularly polarized branches have slightly different dispersion relations.   As a result, if a  linearly polarized transverse wave is launched along the chain, it undergoes Faraday rotation.  For parameters approximating that of a typical metal and  for a field of 2T,  the Faraday rotation  is of order 1$^o$ per ten interparticle spacings, even taking into account single-particle damping. If ${\bf B}$ is perpendicular to the chain, one of the $T$ branches mixes with the $L$ branch to form two elliptically polarized branches.    Our calculations include single-particle damping and can, in principle, be generalized to include radiation damping.   The present work suggests that the dispersion relations of plasmonic waves on chain of nanoparticles can be controlled by immersing the chain in a nematic liquid crystal and varying the director axis, or by applying a magnetic field.

\end{abstract}


\maketitle

\section{Introduction}

The optical properties of small metal particles have been of interest to physicists since the time of Maxwell\cite{maxwell}.  Such particles, if subjected to light of wavelength much larger than their linear dimensions, exhibit optical resonances due to localized electronic excitations known as ``particle'' or `` surface" plasmons.  These plasmons can give rise to characteristic absorption peaks, typically in the near-infrared or the visible, which may play an important role in the optical response of
dilute suspensions of metal particles in a dielectric host\cite{pelton,maier3,solymar}.  

Because of recent advances in sample preparation, it has become possible to study {\it  ordered} arrays of metal particles in a dielectric host\cite{meltzer,maier2,tang,park}.  In one-dimensional ordered arrays of such closely-spaced particles, waves of plasmonic excitations can propagate along the chains, provided that the interparticle spacing is small compared to the wavelength of light\cite{koederink,brong,maier03,park04,plasmonchain,alu06,halas,weber04,simovski05,abajo,jain,crozier}.  In this limit, the electric fields produced by the dipole moment of one nanoparticle induces dipole moments on the neighboring nanoparticles.  The dispersion relations for both transverse ($T$) and longitudinal ($L$) plasmonic waves can then be calculated in the so-called quasistatic approximation\cite{brong,maier03,park04}, in which the curl of the electric field is neglected.   While this approximation neglects some significant coupling between the plasmonic waves and free photons\cite{weber04}, it gives reasonable results over most of the Brillouin zone.  Interest in such plasmonic waves has grown vastly in recent years\cite{park04, plasmonchain}.   

In this paper, we extend the study of propagating plasmonic waves in two ways.  First, we calculate the dispersion relation for such plasmonic waves when the metallic chain is immersed in an anisotropic host, such as a nematic liquid crystal (NLC).  Using a simple approximation, we show that both the $L$ and $T$ waves have modified dispersion relations when the director is parallel to the chain axis.  If the director is perpendicular to that axis, we show that the previously degenerate $T$ branches are split into two separate branches.   Secondly, we consider the effects of a static magnetic field applied either parallel and perpendicular to the chain.  For the parallel case we show that a linearly polarized $T$ wave is rotated as it propagates along the chain.  For a field of 2 tesla and reasonable parameters for the metal, this Faraday rotation may be as large as 1-2$^o$ over ten interparticle spacings.    A perpendicular field mixes together the $L$ branch and one of the $T$ branches, leading to two elliptically polarized branches. These results suggest that either an NLC host or an applied magnetic field could be used as an additional "control knob" to manipulate the properties of the propagating waves in some desired way.

The remainder of this paper is organized as follows.  In the next section, we present the formalism which allows one to calculate the dispersion relations for $L$ and $T$ waves
in the presence of either an anisotropic host or an applied dc magnetic field.  In Section III, we give simple numerical examples, and we follow this
by a brief concluding discussion in Section IV. 

\section{Formalism}

\subsection{Overview}

We consider a chain of identical metal nanoparticles, each a sphere of radius $a$, arranged in a one-dimensional periodic lattice along the $z$ axis.  The n$^{th}$ particle is assumed centered at $(0, 0, nd)$ ($-\infty < n < + \infty$).   The propagation of plasmonic waves along such a chain of nanoparticles has already been considered extensively for the case of isotropic metal particles embedded in a homogeneous, isotropic medium\cite{brong}.  Various works have considered the quasistatic case in which the electric field is assumed to be curl-free; this is roughly applicable when both the radius of the particles and the distance between them are small compared the wavelength of light \cite{brong,maier03,park04}.   The extension of such studies to include radiative corrections, i.\ e., to the case when the electric fields cannot be approximated as curl-free, has also been carried out; these corrections can be very important even in some long-wavelength regimes \cite{weber04}.

Here we consider how the plasmon dispersion relations are modified when the particle chain is immersed in an anisotropic dielectric, such as an NLC, or subjected to an applied dc magnetic field.  For the case of metallic particles immersed in an NLC, we assume that the host medium is a uniaxial dielectric, with principal dielectric constants $\epsilon_\perp$, $\epsilon_\perp$, and $\epsilon_\|$.   For metal particles in the presence of an applied magnetic field, we take the host medium to be vacuum, with dielectric constant unity.  

In the absence of a magnetic field, the medium inside the particles is assumed to have a scalar dielectric function $\epsilon(\omega)$.  If there is a magnetic field along the $z$ axis, the dielectric function of the particles becomes a tensor, whose components may be written 
\begin{eqnarray}\label{eq:dielectric_matrix}
\epsilon_{xx}(\omega) = \epsilon_{yy}(\omega) = \epsilon_{zz}(\omega) = \epsilon(\omega) \nonumber \\
\epsilon_{xy}(\omega) = -\epsilon_{yx}(\omega) = iA(\omega),
\end{eqnarray}
with all other components vanishing \cite{hui}.
In the calculations below, we will assume that the nanoparticles are adequately described by a Drude dielectric function.  In this case, the components of the dielectric tensor take the form  \cite{hui}
\begin{equation}
\epsilon(\omega) = 1 - \frac{\omega_p^2}{\omega(\omega+ i/\tau )} 
\label{eq:epsw}
\end{equation}
and
\begin{equation}
A(\omega) = -\frac{\omega_p^2\tau}{\omega}\frac{(\omega_c\tau)}{(1-i\omega\tau)^2}.
\label{eq:aw}
\end{equation}
Here $\omega_p$ is the plasma frequency, $\tau$ is a relaxation time, and $\omega_c = eB/(mc)$ is the cyclotron frequency, where ${\bf B} = B\hat{z}$ is the
magnetic field, $m$ is the electron mass, and $e$ is its charge.  We will use Gaussian units throughout.  In the limit $\omega\tau \rightarrow \infty$, we may write
\begin{eqnarray} \label{eq:elements_dielectric}
\epsilon(\omega) = 1 - \frac{\omega_p^2}{\omega^2}; \nonumber \\
A(\omega) = \frac{\omega_p^2\omega_c}{\omega^3}.
\end{eqnarray}

\subsection{Uniaxially Anisotropic Host}

 We first assume that the
host has a dielectric tensor $\epsilon_h$ with principal components $\epsilon_\perp$, $\epsilon_\perp$, and $\epsilon_\|$.  Such a form is appropriate, for example, in a nematic liquid crystal below its nematic-to-isotropic
transition.  We begin by writing down the electric field at ${\bf x}$ due to a sphere with a polarization ${\bf P}({\bf x}^\prime)$.  In component form, this field takes the form
(see, e.\ g., Ref \cite{stroud75})
\begin{equation}
E_i({\bf x}) =  -\int{\cal G}_{ji}({\bf x} - {\bf x}^\prime)P_j({\bf x}^\prime)d^3x^\prime,
\label{eq:polfield}
\end{equation}
where repeated indices are summed over, and we use the fact that ${\cal G}_{ji} = {\cal G}_{ij}$. 
In eq.\ (\ref{eq:polfield}), ${\bf  P}({\bf x}^\prime)  = (\epsilon - \epsilon_h){\bf E}({\bf x}^\prime)$ is the polarization of the metallic particle, $\epsilon$ is the dielectric function of the
metal particle,  and 
 ${\bf{\cal G}}$ denotes a 3$\times$3 matrix whose elements are
\begin{equation}
{\cal G}_{ij} = \partial_i^\prime\partial_jG({\bf x}_i- {\bf x}^\prime_j),
\label{eq:gfgrad2}
\end{equation}
where $G({\bf x} - {\bf x}^\prime)$ is a Green's function which satisfies the differential equation (see, e.\ g., Ref.\ \cite{Stroud, stroud75})
\begin{equation}
{\bf \nabla}\cdot\epsilon_h{\bf \nabla}G({\bf x}- {\bf x}^\prime) =-\delta({\bf x} - {\bf x}^\prime).
\end{equation} 
If the host dielectric tensor is diagonal and uniaxial with diagonal components $\epsilon_\perp$, $\epsilon_\perp$ and $\epsilon_\|$, 
which we take for the moment to be parallel to the $x$, $y$, and $z$ axes respectively, this Green's function is given by \cite{stroud75}
\begin{equation}
G({\bf x}- {\bf x}^\prime) = \frac{1}{4\pi\epsilon_{\perp}\epsilon_{\|}^{1/2}}
\left[\frac{(x-x^\prime)^2+(y-y^\prime)^2}{\epsilon_\perp} + \frac{(z-z^\prime)^2}{\epsilon_\|}\right]^{-1/2}.
\label{eq:gfaniso}
\end{equation}
Physically, $-{\cal G}_{ij}({\bf x} - {\bf x}^\prime)$ represents the i$^{th}$ component of electric field at ${\bf x}$ due to a unit point dipole oriented in the j$^{th}$ direction
at ${\bf x}^\prime$, in the presence of the anisotropic host.

The next step is to use this result to obtain a self-consistent equation for plasmonic waves along a chain immersed in an anisotropic host.     To do this, we consider the polarization
of the n$^{th}$ particle, which we write as
${\bf P}_n({\bf x}) = \delta\epsilon{\bf E}_{in,n}({\bf x})$, where ${\bf E}_{in,n}({\bf x})$ is the electric field within the n$^{th}$ particle and $\delta\epsilon=\epsilon-\epsilon_h$.  This field, in turn, is related to the external field acting on the n$^{th}$ particle and
arising from the dipole moments of all the other particles.  We approximate this external field as uniform over the volume of the particle, 
and denote it ${\bf E}_{ext,n}$.  This approximation should be reasonable if the particle radius is not too large compared to the separation between
particles (in practice, an adequate condition is probably $a/d  \leq 1/3$, where $a$ is the particle radius and $d$ the nearest neighbor separation).   Then ${\bf E}_{in, n}$ and
${\bf E}_{ext,n}$ are related by \cite{Stroud}
\begin{equation}
{\bf E}_{in,n} = ({\bf 1} - {\bf \Gamma}\delta\epsilon)^{-1}{\bf E}_{ext,n},
\label{eq:einext}
\end{equation}
where ${\bf \Gamma}$ is a  "depolarization matrix'' defined, for example, in Ref.\ \cite{stroud75}.   ${\bf E}_{ext,n}$  is the
field acting on the n$^{th}$ particle due to the dipoles produced by all the other particles, as given by eq.\ (\ref{eq:polfield}).    
Hence, the dipole moment of the n$^{th}$ particle may be written
\begin{equation}
{\bf p}_n = \frac{4\pi}{3}a^3{\bf P}_{in,n} = \frac{4\pi}{3}a^3{\bf t}{\bf E}_{ext,n}
\label{eq:pinn}
\end{equation}
where 
\begin{equation}
{\bf t} = \delta\epsilon\left({\bf 1}-{\bf \Gamma}\delta\epsilon\right)^{-1}
\label{eq:tmatrix}
\end{equation}
is a ``t-matrix'' describing the scattering properties of the metallic sphere in the surrounding material.   Finally, we make the assumption that the portion of
${\bf E}_{ext,n}$ which comes from particle n$^\prime$ is obtained from eq.\ (\ref{eq:polfield}) as if  the spherical particle $n^\prime$ were a point particle located at the center of
the sphere (this approximation should again be reasonable if $a/d \leq 1/3$).  With this approximation, and
combining eqs.\  (\ref{eq:polfield}),  (\ref{eq:pinn}), and (\ref{eq:tmatrix}), we obtain the following self-consistent
equation for coupled dipole moments:
\begin{equation}
{\bf p}_n = -\frac{4\pi a^3}{3}{\bf t}\sum_{n^\prime \neq n}{\cal G}({\bf x}_n - {\bf x}_{n^\prime}){\bf p}_{n^\prime}.
\label{eq:selfconsist}
\end{equation}

Let us first assume that the principal axis of the anisotropic host coincides with the chain direction, which we take as the $z$ axis.   
In this case, the $L$ and $T$ waves decouple and
can be treated independently, because one of the principal axes of the ${\bf \Gamma}$ tensor coincides with the chain axis.  First, we consider
the longitudinally polarized waves.  To find their dispersion relation, we need to calculate ${\cal G}_{zz}({\bf x}_n - {\bf x}_{n^\prime})$.  From the definition of this quantity,
and from the fact that ${\bf x}_n = nd\hat{z} \equiv z_n\hat{z}$, we can readily show that
${\cal G}_{zz}({\bf x}_n - {\bf x}_{n^\prime})= -\frac{1}{2\pi\epsilon_\perp}\frac{1}{|z_n-z_{n^\prime}|^3}$.
Hence, we obtain the following equations for the $p_{nz}$'s:
\begin{equation}
p_{nz} =-\frac{2}{3\epsilon_{\perp}}a^3\frac{\delta\epsilon_{\|}}{1-\Gamma_\|\delta\epsilon_{\|}}\sum_{n^\prime \neq n}
\frac{p_{n^\prime z}}{|z_n-z_n^\prime|^3}.
\label{eq:coupled}
\end{equation}

For transverse modes, the relevant Green's function takes the form
${\cal G}_{xx}({\bf x}_n - {\bf x}_{n^\prime}) = \frac{\epsilon_{\|}}{4\pi\epsilon_\perp^2}\frac{1}{|z_n-z_{n^\prime}|^3}$.
The resulting equation for the dipole moments  takes the form
\begin{equation}
p_{nx} =\frac{1}{3}\frac{\epsilon_{\|}}{\epsilon_\perp^2} a^3\frac{\delta\epsilon_\perp}{1-\Gamma_\perp\delta\epsilon_\perp}
\sum_{n^\prime \neq n}\frac{p_{n^\prime,x}}{|z_n- z_{n^\prime}|^3}.
\end{equation}
In the isotropic case with a vacuum host, $\epsilon_\|=\epsilon_\perp=1$, and $\Gamma_{xx}=\Gamma_{yy} = \Gamma_{zz} = -1/3$,   The equations for both the parallel and perpendicular
cases reduce to the results obtained in Ref.\ \cite{brong} for both $L$ and $T$ modes, as expected.

For an anisotropic host and only nearest neighbor interactions, the dispersion relation for the $T$ waves is implicitly given by
\begin{equation}
1 = -\frac{2}{3}\frac{a^3}{d^3}\frac{\delta\epsilon_{\perp}}{1-\Gamma_{\perp}\delta\epsilon_{\perp}}\frac{\epsilon_{\|}}{\epsilon_{\perp}^2}\cos kd
\label{eq:twave1}
\end{equation}
and for the $L$ waves by
\begin{equation}
1 = \frac{4}{3}\frac{a^3}{d^3}\frac{\delta\epsilon_{\|}}{1-\Gamma_{\|}\delta\epsilon_{\|}}\frac{1}{\epsilon_{\perp}}\cos kd.
\label{eq:lwave1}
\end{equation}
The forms of $\Gamma_{\perp}$ and $\Gamma_{\|}$ are well known (see, e.\ g., Ref.\ \cite{stroud75}, where they are
denoted $\Gamma_{xx}$ and $\Gamma_{zz}$).  We rewrite them here for convenience:
\begin{eqnarray}
\Gamma_{\|} & = & -\frac{1}{\epsilon_{\|}\lambda}\left[1 - \sqrt{1-\lambda}\frac{\sin^{-1}\sqrt{\lambda}}{\sqrt{\lambda}}\right]
\nonumber \\
\Gamma_{\perp} & = & -\frac{1}{2} \left[\Gamma_{\|} + \frac{1}{\sqrt{\epsilon_{\perp}\epsilon_{\|}}}\frac{\sin^{-1}\sqrt{\lambda}}{\sqrt{\lambda}}\right] 
\label{eq:gammaxz}
\end{eqnarray}
where $\lambda = 1 - \epsilon_{\perp}/\epsilon_\|$.
If we assume that the metallic particle has a Drude dielectric function of the form $\epsilon(\omega) = 1-\omega_p^2/\omega^2$, then 
the dispersion relation for $T$ waves becomes
\begin{equation}
\frac{\omega_t^2(k)}{\omega_p^2} = \frac{T(k)}{(1-\epsilon_{\perp})T(k)-1},
\label{eq:tdisp1}
\end{equation}
where
\begin{equation}
T(k)= \Gamma_{\perp} - \frac{2}{3}\frac{a^3}{d^3}\frac{\epsilon_{\|}}{\epsilon_{\perp}^2}\cos(kd),
\end{equation}
and that of the $L$ waves is
\begin{equation}
\frac{\omega_\ell^2(k)}{\omega_p^2} = \frac{L(k)}{(1-\epsilon_{\|})L(k) - 1},
\label{eq:ldisp1}
\end{equation}
where
\begin{equation}
L(k) = \Gamma_{\|} + \frac{4}{3}\frac{a^3}{d^3}\frac{1}{\epsilon_{\perp}}\cos(kd).
\end{equation}

Eqs.\  (\ref{eq:tdisp1}) and (\ref{eq:ldisp1}) neglect damping of the waves due to dissipation within the metallic particles.  To include this effect, one can
simply solve eq.\  (\ref{eq:twave1}) or (\ref{eq:lwave1}) for $k(\omega)$, using the Drude function with a finite $\tau$.   The resulting $k(\omega)$ will be complex
in both cases; the inverse of the imaginary part of $k(\omega)$ will give the exponential decay length of the $T$ or $L$ wave along the chain.  

Now, let us repeat this calculation but with the principal axis of the liquid crystalline host parallel to the $x$ axis, while the chain itself again lies along
the $z$ axis.   The self-consistent equation for the dipole moments again takes the form  (\ref{eq:selfconsist}), but the diagonal elements of ${\cal G}$ 
are given by ${\cal G}_{ii}({\bf x}, {\bf x}^\prime)  = \partial_i^\prime\partial_iG({\bf x}- {\bf x}^\prime)$, where now
\begin{equation}
G({\bf x}- {\bf x}^\prime) = \frac{1}{4\pi\epsilon_\perp\epsilon_\|^{1/2}}\left[\frac{(x-x^\prime)^2}{\epsilon_\|}
+ \frac{(y-y^\prime)^2 + (z-z^\prime)^2}{\epsilon_\perp}\right]^{-1/2}.
\end{equation}
For the case of interest, ${\bf x} = nd{\bf \hat{z}} \equiv z_n{\bf \hat{z}}$, ${\bf x}^\prime = n^\prime d{\bf \hat{z}} \equiv z_{n^\prime}{\bf \hat{z}}$, and one finds
that ${\cal G}_{xx}({\bf x} - {\bf x}^\prime)  =  \frac{1}{4\pi}\frac{\epsilon_\perp^{1/2}}{\epsilon_\|^{3/2}}\frac{1}{|z_n-z_{n^\prime}|^3}$,
 ${\cal G}_{yy}({\bf x} - {\bf x}^\prime)  =    \frac{1}{4\pi}\frac{1}{\epsilon_\perp^{1/2}\epsilon_\|^{1/2}}\frac{1}{|z_n-z_{n^\prime}|^3}$
and ${\cal G}_{zz}({\bf x}-{\bf x}^\prime)  =  -\frac{1}{2\pi}\frac{1}{\epsilon_\perp^{1/2}\epsilon_\|^{1/2}}\frac{1}{|z_n-z_{n^\prime}|^3}$.
The self-consistency condition determining the relation between $\omega$ and $k$ can be written out, for all three polarizations, including
only nearest neighbor dipole-dipole interactions, in the form $1 = -8\pi \frac{a^3}{d^3}\frac{\delta\epsilon_{ii}}{1-\Gamma_{ii}\delta\epsilon_{ii}}{\cal G}_{ii}(d)\cos kd$.
Substituting in the values of ${\cal G}_{ii}$ for the three cases, we obtain
\begin{eqnarray}
1 & = & -\frac{2a^3}{3d^3}\frac{\delta\epsilon_{xx}}{1-\Gamma_{xx}\delta\epsilon_{xx}}\frac{\epsilon_\perp^{1/2}}{\epsilon_\|^{3/2}}\cos kd \nonumber \\
1 & = & -\frac{2a^3}{3d^3}\frac{\delta\epsilon_{yy}}{1-\Gamma_{yy}\delta\epsilon_{yy}}\frac{1}{\epsilon_\|^{1/2}\epsilon_\perp^{1/2}}\cos kd  \nonumber \\
1 & = & \frac{4a^3}{3d^3}\frac{\delta\epsilon_{zz}}{1-\Gamma_{zz}\delta\epsilon_{zz}}\frac{1}{\epsilon_\perp^{1/2}\epsilon_\|^{1/2}} \cos kd.
\end{eqnarray}
Here $\Gamma_{xx} = \Gamma_\|$, $\Gamma_{yy} = \Gamma_{zz} = \Gamma_\perp$, where $\Gamma_\|$ and $\Gamma_\perp$ are given by eqs.\ (\ref{eq:gammaxz}).  Similarly, $\delta\epsilon_{xx} = \epsilon(\omega) - \epsilon_\|$, while $\delta\epsilon_{yy}=\delta\epsilon_{zz}=\epsilon(\omega)-\epsilon_\perp$.  These equations can again be solved for $k(\omega)$ in the three cases, with or without a finite $\tau$, leading to dispersion relations with or without single-particle damping.

\subsection{Chain of Metallic Nanospheres in an External Magnetic Field}

Next, we turn to a chain of metallic nanoparticles in an external magnetic field, which we initially take to be parallel to the $z$ axis.   For such a system, we assume that the metal dielectric tensor is of the form (\ref{eq:dielectric_matrix}), (\ref{eq:epsw}) and (\ref{eq:aw}).   The cases of a dilute suspension of metallic particles\cite{hui}, or of a random composite of ferromagnetic and non-ferromagnetic particles\cite{xia}, have been treated previously.

Once again, we take the chain of particles to lie along the $z$ axis, with the n$^{th}$ particle centered at $z_n= nd$.   The self-consistent equation for the dipole moments is still eq.\  (\ref{eq:selfconsist}), but now the elements of both ${\cal G}$ and ${\bf \Gamma}$ are different from the case of an NLC host.   For a chain of particles parallel to the $z$ axis, ${\cal G}$ is diagonal, with non-zero elements
${\cal G}_{xx}({\bf x} -{\bf x}^\prime) = {\cal G}_{yy}({\bf x}- {\bf x}^\prime) = \frac{1}{4\pi|z_n-z_{n^\prime}|^3}$,
${\cal G}_{zz}({\bf x} -{\bf x}^\prime)  =  -\frac{1}{2\pi|z_n-z_{n^\prime}|^3}$,  where $z_n = nd$ and
we have assumed that the host has a dielectric constant equal to unity.   The tensor ${\bf \Gamma}$ is also diagonal, with nonzero elements  $\Gamma_{ii} = -1/3$, i = 1, 2, 3.  The quantity $\delta\epsilon = \epsilon(\omega) -  1$, where $\epsilon(\omega)$ is now the dielectric tensor of the metallic particle. Using eq.\ (\ref{eq:dielectric_matrix}) to evaluate this tensor, we obtain the following result for the tensor ${\bf t} \equiv \delta\epsilon[1-{\bf \Gamma}\delta\epsilon]^{-1}$, to first order in the quantity $A(\omega)$, which is assumed to be small:
\begin{eqnarray}
t_{zz} & = & \delta\epsilon_{zz}(1 - \Gamma_{zz}\delta\epsilon_{zz})^{-1} \nonumber \\
t_{xx}=t_{yy} & = & \delta\epsilon_{xx}(1-\Gamma_{xx}\delta\epsilon_{xx})^{-1} \nonumber \\
t_{xy} = -t_{yx} & = & \delta\epsilon_{xx}\Gamma_{xx}\delta\epsilon_{xy}(1-\Gamma_{xx}\delta\epsilon_{xx})^{-2}.
\end{eqnarray}
Using these expressions, we can now write out the self-consistent linear equations for the oscillating dipole moments and obtain dispersion relations for the modes. Once again, the longitudinal and transverse modes decouple.   For the longitudinal modes, the self-consistent equation simplifies to
\begin{equation}
p_{nz} = 2a^3\frac{\epsilon(\omega)-1}{\epsilon(\omega)+2}\sum_{n^\prime \neq n}\frac{p_{n^\prime z}}{|z_n-z_n^\prime|^3}.
\label{eq:pzmag}
\end{equation}
This is the same as the equation for the longitudinal modes in the absence of a magnetic field and gives the same dispersion relation.  For the transverse modes, the $x$ and $y$ components
of the polarization are now coupled, and satisfy the equations [writing ${\cal G}_{ij}({\bf x} - {\bf x}^\prime) = {\cal G}_{ij}(z_n - z_{n^\prime})$]
\begin{eqnarray}
p_{nx} & = & -\frac{4\pi}{3}a^3 \sum_{n^\prime \neq n} \left[t_{xx}{\cal G}_{xx}(z_n-z_{n^\prime})p_{n^\prime, x} + t_{xy}{\cal G}_{yy}(z_n-z_{n^\prime})p_{n^\prime,y}\right] \nonumber \\ 
p_{ny} & = & -\frac{4\pi}{3}a^3 \sum_{n^\prime \neq n} \left[t_{yx}{\cal G}_{xx}(z_n-z_{n^\prime})p_{n^\prime, x} + t_{yy}{\cal G}_{yy}(z_n-z_{n^\prime})p_{n^\prime,y}\right]
\label{eq:pxymag}
\end{eqnarray}
We can simplify these equations using the fact that ${\cal G}_{xx}(z_n-z_{n^\prime})={\cal G}_{yy}(z_n-z_{n^\prime})$, $t_{xx}=t_{yy}$, and $t_{xy} = t_{yx}$ to obtain
\begin{equation}
p_{n,+} = -\frac{4\pi}{3}a^3t_{\perp,-}\sum_{n^\prime \neq n}{\cal G}(z_n-z_{n^\prime})p_{n^\prime,+}
\label{eq:pxymag1}
\end{equation}
and
\begin{equation}
p_{n,-} = -\frac{4\pi}{3}a^3t_{\perp,+}\sum_{n^\prime \neq n}{\cal G}(z_n-z_{n^\prime})p_{n^\prime,-},
\label{eq:pxymag2}
\end{equation}
where $t_{\perp,\pm} = t_{xx} \pm it_{xy}$ and $p_{n,\pm} = p_{nx} \pm ip_{ny}$.  Thus, the equations for left- and right-circularly polarized waves are decoupled.

To obtain explicit dispersion relations for left- and right-circularly polarized waves, we assume, as before, that $p_{n,\pm} = p_\pm\exp(iknd-i\omega t)$, and substitute the
known forms for the quantities $t_{xx}$, $t_{xy}$, and $G_{xx}(z_n-z_{n^\prime})$, with the following result:    
\begin{equation}
1 = -2\frac{a^3}{d^3}\left[\frac{\epsilon(\omega)-1}{\epsilon(\omega)+2} \pm \frac{3A(\omega)}{\left[\epsilon(\omega)+2\right]^2}\right]\sum_{n=1}^\infty\frac{cos(nk_\pm d)}{n^3}.
\label{eq:dispers1}
\end{equation}
 In the special case where we include dipolar interactions only
between nearest neighbors, this relation becomes
\begin{equation}
1 = -2\frac{a^3}{d^3}\left[\frac{\epsilon(\omega)-1}{\epsilon(\omega)+2}\pm \frac{3A(\omega)}{\left[\epsilon(\omega)+2\right]^2}\right]\cos(k_\pm d).
\label{eq:dispers2}
\end{equation}
Since the frequency-dependence of both $\epsilon(\omega)$ and $A(\omega)$ is assumed known, these equations represent implicit relations between $\omega$ and $k_\pm$ for these transverse waves.

By solving for $k_\pm(\omega)$ in eq.\ (\ref{eq:dispers1}) or (\ref{eq:dispers2}), one finds that left and right circularly polarized transverse waves propagating along the nanoparticle chain have slightly different wave vectors $k_+$ and $k_-$ for the same frequency $\omega$.   Since a linearly polarized wave is composed of an equal fraction of right and left circularly polarized waves,  this behavior corresponds to a {\it rotation} of the plane of polarization of a linearly polarized wave, as it propagates down the chain, and is analogous to the usual Faraday effect in a {\it bulk} dielectric.   The angle of rotation per unit chain length may be written
\begin{equation}
\label{eq:angle_single}
\theta(\omega) = \frac{1}{2} \left[k_+(\omega) - k_-(\omega)\right].
\end{equation}

In the absence of damping, $\theta$ is real.  If $\tau$ is finite, the electrons in each metal particle will experience damping within each particle, leading to an exponential
decay of the plasmons propagating along the chain. This damping is automatically  included in the above formalism, and can be seen most easily
if only nearest neighbor coupling is included. 
The quantity
\begin{equation}
\theta(\omega) = \theta_1(\omega) + i\theta_2(\omega) 
\label{eq:thetatau}
\end{equation}
is then the {\it complex} angle of rotation per unit length of a linearly polarized wave propagating along the chain of metal particles.   By analogy with the interpretation of a complex $\theta$ in a homogeneous bulk material, Re$\theta(\omega)$ represents the angle of rotation of a linearly polarized wave (per unit length of chain), and  Im$\theta(\omega)$ as the corresponding Faraday ellipticity - i.\ e., the degree to which the initially linearly polarized wave becomes elliptically polarized as it propagates along the chain.  

For a magnetic field perpendicular to the chain (let us say, along the $x$ axis), the elements of the matrix ${\bf t}$ become
\begin{eqnarray} 
t_{xx} =t_{yy}=t_{zz} & = & 3[\epsilon(\omega)-1]/[\epsilon(\omega)+2]  \nonumber \\
t_{yz} = -t_{zy} &= &-3iA(w)[\epsilon(\omega)-1]/[\epsilon(\omega)+2]^2, 
\end{eqnarray}
with other elements equal to zero. The transverse waves polarized parallel to the $x$ axis are unaffected by the magnetic field, and are described by the equations
\begin{equation}
p_{nx} = -\frac{4\pi}{3}a^3t_{xx}\sum_{n^\prime\neq n}{\cal G}_{xx}(z_n-z_{n^\prime})p_{n^\prime,x}.
\end{equation}
The $y$ and $z$ polarized waves, with components $p_{ny}$ and $p_{nz}$, are coupled, however, and satisfy
\begin{eqnarray}
p_{ny} & = & -\frac{4\pi}{3}a^3\sum_{n^\prime\neq n}\left[t_{yy}{\cal G}_{yy}(z_n- z_{n^\prime})p_{n^\prime, y} + t_{yz}{\cal G}_{zz}(z_n-z_{n^\prime})p_{n^\prime, z}\right] \nonumber \\
p_{nz} & = & -\frac{4\pi}{3}a^3\sum_{n^\prime\neq n}\left[t_{zy}{\cal G}_{yy}(z_n-z_{n^\prime})p_{n^\prime, y} + t_{zz}{\cal G}_{zz}(z_n-z_{n^\prime})p_{n^\prime, z}\right]
\end{eqnarray}
The tensor ${\cal G}$ is still diagonal, with the same nonzero elements as in the case of ${\bf B}$ parallel to the chain.
Assuming propagating waves of the form $p_{ny} = p_{0y}\exp(inkd-i\omega t)$, $p_{nz} = p_{0z}\exp(inkd-i\omega t)$,
we find that the amplitudes $p_{0y}$ and $p_{0z}$ satisfy the equations
\begin{eqnarray} p_{0y} & = & -\frac{4\pi}{3}a^3\sum_{n^\prime \neq 0}\left[t_{yy}{\cal G}_{yy}(-z_{n^\prime})p_{0y} +
t_{yz}{\cal G}_{zz}(-z_{n^\prime})p_{0z}\right]exp(ikn^\prime d)  \nonumber  \\
p_{0z} & = & -\frac{4\pi}{3}a^3\sum_{n^\prime \neq 0}\left[-t_{yz}{\cal G}_{yy}(-z_{n^\prime})p_{0y}+t_{zz}{\cal G}_{zz}(-z_{n^\prime})p_{0z}\right]\exp(ikn^\prime d).
\label{eq:rotyz}
\end{eqnarray}
In the special case where only nearest neighbor interactions are included, these equations simplify to 
\begin{eqnarray}
p_{0y} & = &-\frac{8\pi}{3}a^3\left[t_{yy}{\cal G}_{yy}(d)p_{0y}+t_{yz}{\cal G}_{zz}(d)p_{0z}\right]\cos(kd) \nonumber \\
p_{0z} &= & -\frac{8\pi}{3}a^3\left[-t_{yz}{\cal G}_{yy}(d)p_{0y} + t_{zz}{\cal G}_{zz}(d)p_{0z}\right]\cos(kd).
\label{eq:rotyz1}
\end{eqnarray}
Substituting in the explicit forms of ${\cal G}_{yy}$ and ${\cal G}_{zz}$, we find that these equations take the form
\begin{eqnarray}
p_{0y} & = & \frac{a^3}{d^3}\left[-\frac{2}{3}t_{yy}p_{0y} + \frac{4}{3}t_{yz}p_{0z}\right]\cos kd \nonumber \\
p_{0z} & = & -\frac{a^3}{d^3}\left[\frac{2}{3}t_{yz}p_{0y} + \frac{4}{3}t_{zz}p_{0z}\right]\cos kd.
\label{eq:rotyz2}
\end{eqnarray}

If we solve the pair of equations (\ref{eq:rotyz}) or (\ref{eq:rotyz2}) for $p_{0y}$ and $p_{0z}$ for a given value of $k$, we obtain a nonzero solutions only if the determinant of the matrix of coefficients vanishes. 
For a given real frequency $\omega$, there will, in general, be two solutions for $k(\omega)$ which decay in the $+z$ direction.   These correspond to two branches of propagating plasmon (or plasmon polariton) waves, with dispersion relations which we may write $k_1(\omega)$ and $k_2(\omega)$.  The frequency dependence appears because both $t_{yz}$ and $t_{yy}$ depend on $\omega$ [through $\epsilon(\omega)$ and $A(\omega)$].  The corresponding solutions $(p_{0y}$, $p_{0z})$ are no longer linearly polarized but will instead be elliptically polarized.   However, unlike the case where the magnetic field is parallel to the  $z$ axis, the waves are not circularly polarized, and the two solutions are non-degenerate.   Because $A(\omega)$ is usually small, the ellipse has a high eccentricity, and the change in propagation characteristics due to the magnetic field will usually be small for this magnetic field direction.

\section{Numerical Illustrations}

As a first numerical example, we calculate the plasmon dispersion relations for a chain of spherical Drude metal particles immersed in an NLC.   We consider two cases: liquid crystal director parallel and perpendicular to the chain axis, which we take as the $z$ axis.   For $\epsilon_\|$ and $\epsilon_\perp$, we take the values used in Ref.\ \cite{Park}.  These
are taken from experiments described in Ref.\ \cite{muller}, which were carried out on the NLC known as E7.   For comparison, we also show the corresponding dispersion relations for an isotropic host of dielectric constant which is arbitrarily taken as $\frac{1}{3}\epsilon_\| + \frac{2}{3}\epsilon_\perp = 2.5611$.
The results of these calculations are shown in Figs.\ 1  and 2 in the absence of damping ($\tau \rightarrow \infty$ in the Drude expression).   As can be seen, both the $L$ and $T$ dispersion relations are significantly altered when the host is a nematic liquid
crystal rather than an isotropic dielectric; in particular, the widths of the $L$ and $T$ bands are changed.   When the director is perpendicular to the chain axis, the two $T$ branches are split when the host is an NLC, whereas they are degenerate for an isotropic host, or an NLC host with director parallel to the chain.     

Next, we turn to the effects of an applied magnetic field on these dispersion relations for the case ${\bf B} \| {\bf \hat{z}}$.  The $L$ waves are unaffected by a magnetic field, but the $T$ waves are split into left- and right-circularly polarized waves.   To illustrate the predictions of our simple expressions, we again take
$a/d = 1/3$, and we assume a magnetic field such that the ratio $\omega_c/\omega_p = 3.5\times 10^{-5}$.   For a typical metallic plasma frequency of $\sim 10^{16}$ sec$^{-1}$,
this ratio would correspond to a magnetic induction $B \sim 2$ T.   We consider both the undamped case ($\tau \rightarrow \infty$)
and the damped case ($\omega_p\tau = 100$).    Using these parameters, the dispersion relations for the two circular polarizations are shown in Fig.\ 3 both with and without
single-particle damping.   The splitting between the two circularly polarized $T$ waves is not visible on the scale of the figure.  

We also plot the corresponding rotation angles $\theta d$ for a distance equal to one interparticle spacing in Fig.\ 4.    When there is no damping,  $\theta$ diverges near the edge of the $T$ bands, but this divergence disappears for a finite $\tau$ (e.\ g.\ $\omega_p\tau = 100$, as shown in in Fig.\ 4).  In this case,  Re$\theta(\omega) d$ never exceeds about $0.001$ rad per
interparticle separation, corresponding to a rotation of less than 0.1$^o$ over this distance.   Im$\theta(\omega)d$ is also small, showing that a linear incident wave  acquires little ellipticity over such distances.   Over a distance of thirty or so interparticle separations, a linearly polarized transverse wave would typically rotate by only 1-3$^o$.  Since theory and experiment suggests that the wave {\it intensity} typically has an exponential decay length  of no more than around 20 interparticle spacings \cite{book}, the likely Faraday rotation of such a wave in practice will probably not exceed a degree or two, at most, even for a field as large as 2 T.  Thus, while the rotation found here  is likely to be measurable, it may not be large, at least for this simple chain geometry with one  particle per unit cell.    The present expressions also indicate that $\theta$ is very nearly linear in B, so a larger rotation could be attained by increasing B. 

As can be seen, $\theta$ depends strongly on frequency in the case of zero damping, but less so at finite damping.   In the (unrealistic) no-damping case, at the very top and the very bottom of the plasmonic band, only one of the two circularly polarized waves can propagate down the chain.  Because this filtering occurs only over a very narrow frequency range (of order $\omega_c/\omega_p$), and because this calculation assumes no damping, it would be quite difficult to detect a region where only one of the two circularly polarized waves can propagate.

Finally, we mention the case where ${\bf B} \perp {\bf \hat{z}}$ for the same parameters as in the parallel case.   The effect of a finite B  causes two non-degenerate dispersion relations (one an $L$ and the other a $T$ wave)  for the perpendicular case to become mixed.   We have not computed the rotation angles for this perpendicular case, but we expect that they would be similar in magnitude to the values shown in Fig.\ 4.

\section{Discussion}

The calculations and formalism presented in the previous section leave out several effects which may be at least quantitatively important.    First, in our numerical calculations, but not in the
formalism, we have included only nearest neighbor dipolar coupling.   Inclusion of further neighbors will quantitatively alter the dispersion relations in all cases considered, as well as the
Faraday rotation angle when there is an applied magnetic field, but these effects should not be very large, as is already suggested by the early calculations in
Ref.\ \cite{brong} for an isotropic host.   Another possible effect will appear when $a/d$ is significantly greater than $1/3$, namely, the emergence of quadrupolar and higher quasistatic bands\cite{park04}.  These will mix with the dipolar band and alter its shape.  For the separations we consider, this multipolar effect should be small.  Also, even if $a > d/3$, the plasmon dispersion relations will still be altered by an NLC host or by an applied magnetic field in the manner described here.

The present treatment also omits radiative damping.  In the absence of a magnetic field, such damping is known to be small but non-zero in the long-wavelength limit, but it becomes more significant when the particle radius is a substantial fraction of a wavelength.  Even at long wavelengths, radiative damping can be very important at certain characteristic values
of the wave vector\cite{weber04}.  We have not, as yet, extended the present approach to include such radiative effects.  We expect that, just as for an isotropic host in the absence of a 
magnetic field, radiative effects will further damp the propagating plasmons in the geometries we consider, but will not qualitatively change the
effects we have described.

For the case when the host is an NLC, the present work oversimplifies the treatment of the NLC host by assuming that the director field is {\it uniform}, i.\ e., position-independent.   In reality, the director is almost certain to be
modified close to the metal nanoparticle surface, i.\ e., to become nonuniform, as has been pointed out by many authors\cite{lubensky}.  The effects of such complications on the optical properties of a single metallic particle immersed in an NLC have been treated, for example, in Ref.\ \cite{park05}, and similar approaches might be possible for the present problem also.

To summarize, we have shown that the dispersion relations for plasmonic waves propagating along a chain of closely spaced nanoparticles of Drude metal are strongly affected by external perturbations.  First, if  the host is a uniaxially anisotropic dielectric (such as a nematic liquid crystal), the dispersion relations of both $L$ and $T$ modes are significantly modified, compared to those of an isotropic host, and if the director axis of the NLC is perpendicular to the chain, the two degenerate transverse branches are split.  Secondly, if the chain is subjected to an applied magnetic field parallel to the axis, the $T$ waves undergo a small but measurable Faraday rotation, and also acquire a slight ellipticity.
A similar ellipticity  develops if the magnetic field is perpendicular to the chain, but its effect will likely be more difficult to observe, because the magnetic field couples two non-degenerate branches of the dispersion relation.   All  these effects show that the propagation of such plasmonic waves can be tuned, by either a liquid crystalline host or a magnetic field, so as to change the frequency band  where wave propagation can occur, or the polarization of these waves.   This
control may be valuable in developing devices using plasmonic waves in future optical circuit design.

\section{Acknowledgments}

This work was supported by the Center for Emerging Materials at The Ohio State University, an NSF MRSEC (Grant No.\ DMR0820414).

\newpage

\begin{figure}[ht]
\includegraphics[scale=1.0]{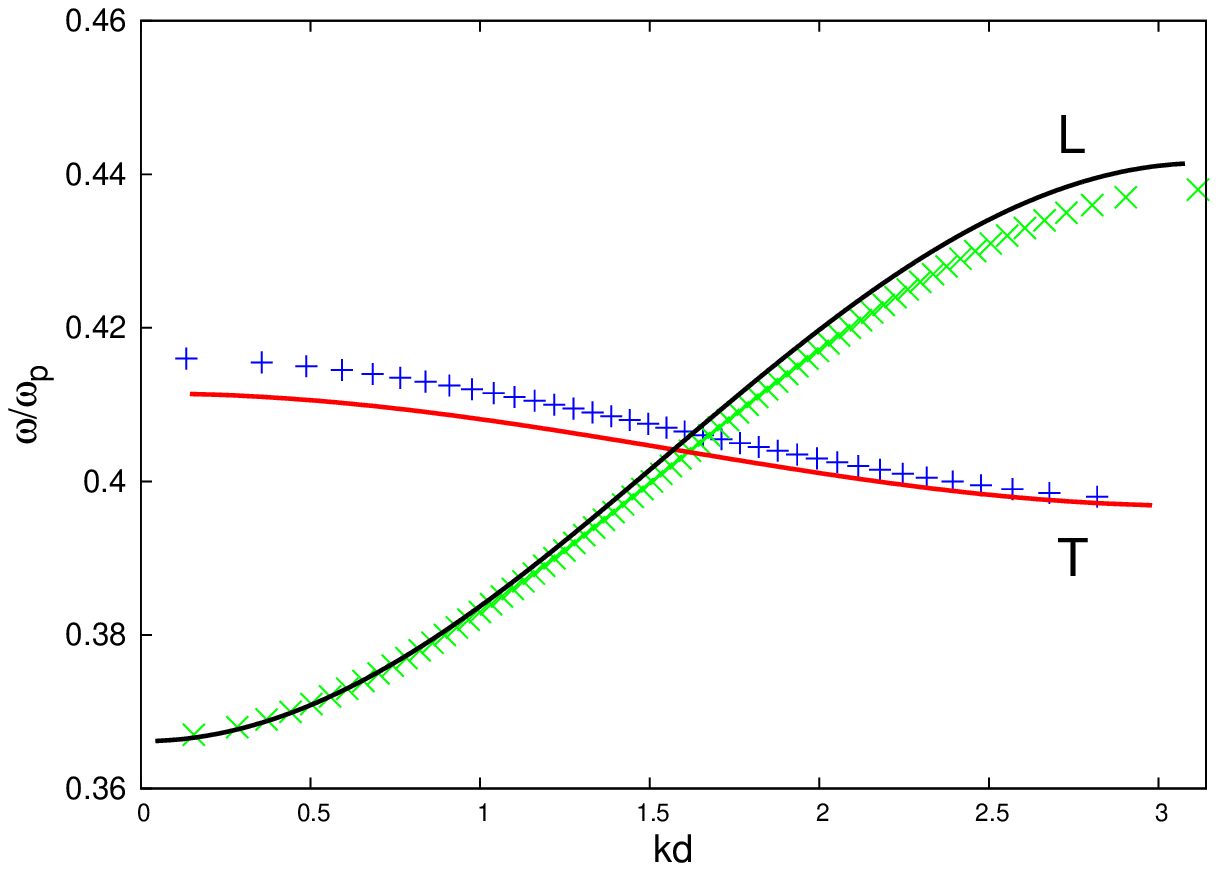}
\caption{Calculated dispersion relations $\omega(k)$ for plasmon waves along a chain of metallic nanoparticles, in the presence of an NLC host.   We plot $\omega/\omega_p$, where
$\omega_p$ is the plasma frequency, as a function of $kd$, where $d$ is the distance between sphere centers.  Green and blue (x's and +'s):  $L$ and $T$ modes for a chain embedded in an NLC with director parallel to the chain.   The NLC is assumed to have principal dielectric tensor elements $\epsilon_\| = 3.0625$ and $\epsilon_\perp = 2.3104$  parallel and perpendicular to the director, corresponding to the material known as E7.   In this and all subsequent plots $a/d = 1/3 $, where $a$ is the metallic sphere radius. 
Also shown are the corresponding $L$ and $T$ dispersion relations (black and red solid lines, respectively) when the host is isotropic with dielectric constant $\epsilon_h = 2.5611 = \frac{1}{3}\epsilon_\| + \frac{2}{3}\epsilon_\perp$.} 

\label{figure1}
\end{figure}

\newpage

\begin{figure}[ht]
\includegraphics[scale=1.0]{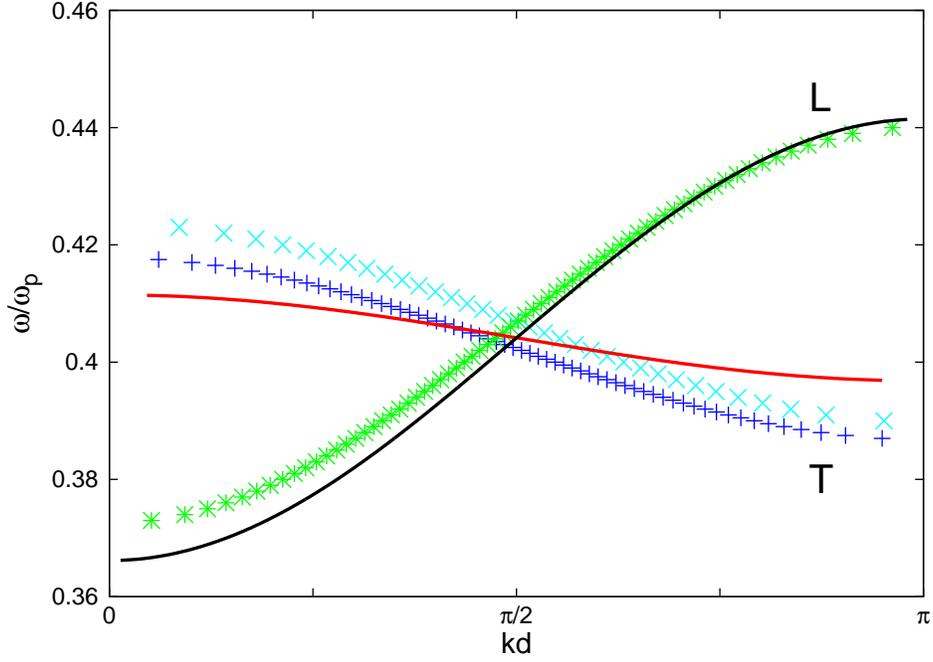}
\caption{Same as Fig.\ 1 except that the director of the NLC is perpendicular to the chain of metal nanoparticles.  The frequencies of
the $L$ modes (asterisks, in green)
and $T$ modes (+'s and x's, shown in dark and light blue), divided by the plasma frequency $\omega_p$,  are plotted versus $kd$.  The NLC has the same dielectric tensor elements as in Fig.\ 1.   Also shown are the corresponding $L$ (solid black) and $T$ (solid red) branches for an isotropic host with $\epsilon_h = 2.5611$.   Note that the $T$ branches which were 
degenerate in Fig.\ 1 are split into two branches in this NLC geometry.}
\label{figure2}
\end{figure}

\newpage

\begin{figure}[ht]
\includegraphics[scale =1.0]{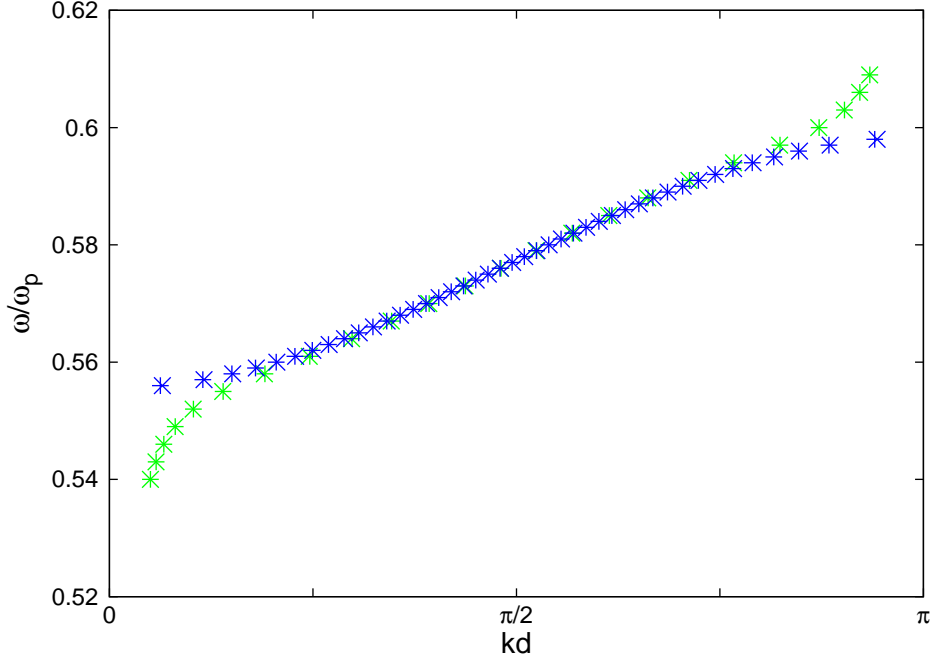} 
\caption{Blue symbols (+ and x symbols): the dispersion relations for left and right circularly polarized $T$  plasmon waves propagating along a chain of metallic nanoparticles, assuming no damping and a magnetic field parallel to the chain.  We assume also that $\omega_c/\omega_p = 3.5 \times  10^{-5}$, where $\omega_c = eB/mc$ is the cyclotron frequency.   Green + and x symbols: same dispersion relations, but assuming single-particle damping corresponding to $\omega_p\tau = 100$.   In both cases, the splitting between left and right circularly polarized waves is not visible on the scale of the figure, so that for both undamped and damped dispersion relations there appear to be single sets of asterisks.   For $\omega_p = 1.\times 10^{16}$ sec$^{-1}$, the chosen
$\omega_c/\omega_p$ corresponds to about 2 T. }  
\label{figure3}
\end{figure}

\newpage

\vspace{1.0in}

\begin{figure}[ht]
\includegraphics[scale =1.0]{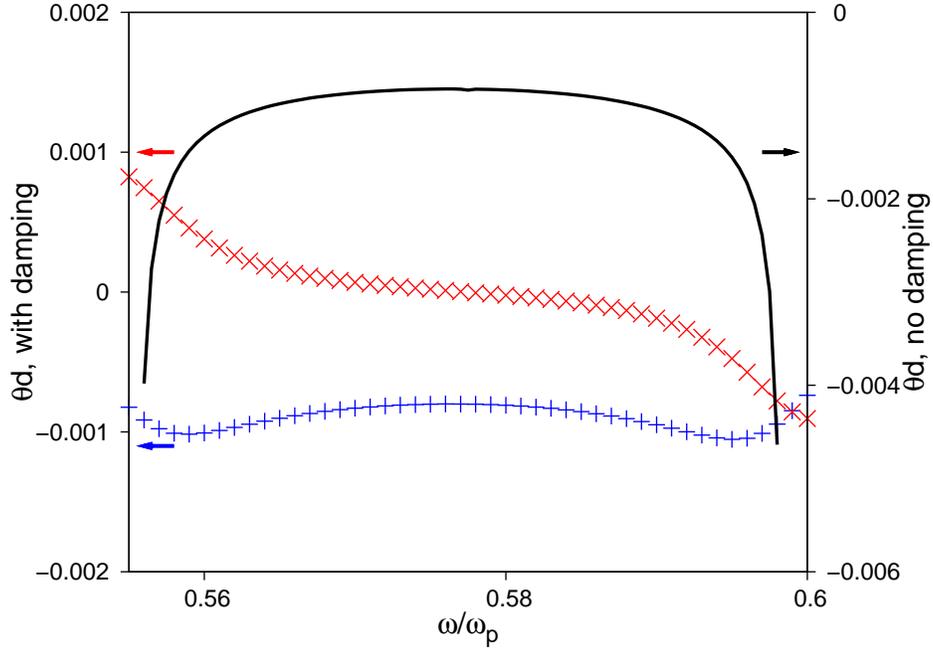} 
\caption{Solid curve: Faraday rotation angle $\theta d$ per interparticle spacing as a function of frequency, assuming $\tau\rightarrow \infty$, and 
$\omega_c/\omega_p = 3.5\times 10^{-5}$.    Blue and red (+ and x) symbols: Real  and imaginary parts of the Faraday rotation angle per interparticle spacing, $\theta d$, assuming $\omega_p\tau = 100$ and
the ratio $\omega_c/\omega_p = 3.5 \times 10^{-5}$.  Note that there are two different vertical scales for undamped and damped rotation angles.}
\label{figure4}.
\end{figure}

\end{document}